\documentstyle[12pt]{article}

\newcommand{\be}{\begin{equation}}
\newcommand{\ee}{\end{equation}}

\catcode `\@=11
\@addtoreset{equation}{section}

\def\section{\@startsection {section}{1}{\z@}{-3.5ex plus -1ex minus
     -.2ex}{2.3ex plus .2ex}{\normalsize\bf}}
\def\subsection{\@startsection{subsection}{2}{\z@}{-3.25ex plus -1ex 
minus
 -.2ex}{1.5ex plus .2ex}{\normalsize\bf}}
 
\catcode `\@=12

\title{\bf\huge Schwarzschild black hole with \\
global monopole charge}
\author{Naresh Dadhich \thanks{E-mail : nkd@iucaa.ernet.in} \\
{\sl Inter-University Centre for Astronomy \& Astrophysics,}\\
{\sl Post Bag 4, Ganeshkhind, Pune - 411 007, India.} \\
K. Narayan \thanks{E-mail :  narayan@niharika.phy.iitb.ernet.in} and
U. A. Yajnik \thanks{E-mail :  yajnik@niharika.phy.iitb.ernet.in}\\
{\sl Department of Physics,}\\
{\sl Indian Institute of Technology Bombay,} \\
{\sl Mumbai 400 076.}
} 
  
\date{}
\begin{document}

\maketitle

\begin{abstract}
We derive the metric for a Schwarzschild black hole with global
monopole charge by relaxing asymptotic flatness of the Schwarzschild
field. We then study the effect of global monopole charge on particle
orbits and the Hawking radiation. It turns out that existence, boundedness 
and stability of circular orbits scale up by $(1-8 \pi\eta^2)^{-1}$, and the
perihelion shift and the light bending by $(1-8 \pi\eta^2)^{-3/2}$,
while the Hawking temperature scales down by $(1 - 8 \pi \eta^2)^2$
the Schwarzschild values. Here $\eta$ is the global charge.
\end{abstract}


\vfill
\begin{flushright} IITB-TP-97101 \end{flushright}
\vfill

\newpage

\section{ Introduction}

The Schwarzschild solution is the
unique spherically symmetric solution
of the vacuum Einstein equation.
It represents a static black hole. It
turns out [1] that a global monopole charge accompanies 
spontaneous
breaking of global $0(3)$ symmetry into $U(1)$ in phase
transitions in
the Universe. Putting on global charge to the 
Schwarzschild black hole
will amount to breaking the vacuum and asymptotic 
flatness of the 
Schwarzschild spacetime. It still represents a 
localized object with
a horizon. Our main aim in this paper is to study the 
effect of global
monopole charge on particle orbits and the Hawking 
radiation.\\

The spacetime of pure global monopole charge, when the Schwarzschild
mass vanishes, can be regarded in some sense to be ``minimally'' curved
for gravitational charge density, $4 \pi \rho_c = R_{ik} u^i u^k,
u^i u_i = 1$, vanishes. That means the relativistic (active gravitational)
mass is zero for the spacetime [2]. All radial trajectories will 
always remain parallel. Thus introduction of global monopole
charge does not significantly alter the nature of the Schwarzschild
field. We wish to investigate what effect does its introduction
entail on physically measurable quantities.
Recently a similar investigation has been carried out for a global
monopole in the Kaluza-Klein spacetime [3].\\

In  Sec. 2  we  shall  briefly  outline  the 
derivation  of the metric for a static black hole with  a  global 
monopole  charge   followed  in  Sec. 3  by  field   theoretic 
considerations.  The effect of global charge on  particle  orbits 
and  the Hawking radiation process will be investigated  in  Secs. 
4 and 5. We conclude with a discussion in Sec. 6. \\

\section{ The Metric}

Let us write the general spherically symmetric 
metric \\
\be
ds^2 = B dt^2 - A dr^2 - r^2(d \theta^2 + \sin^2 \theta d \varphi^2) 
\ee

\noindent where $A$ and $B$  are  functions  of $r$ and $ t$. \\

\noindent Now the equations $R_{01} = 0$ and $R^0_0 = R^1_1$ will
imply $B = A^{-1} = 1 + 2 \phi$, where $\phi = \phi(r)$ [4]. Note that
no boundary condition has been used. With this 

\be 
R^0_0 = R^1_1 = -\bigtriangledown^2 \phi 
\ee

\noindent and \\

\be R^2_2 = R^3_3 = -\frac{2}{r^2} (r \phi)^{\prime} 
\ee

\noindent where a dash denotes derivative w.r.t. $r$. Then $R^0_0 = 0$
gives the good old Laplace equation which  has the  well-
known general solution 

\be
\phi = k - \frac{M}{r}.
\ee

\noindent Now $R^2_2 = 0$   will determine $k = 0$ and  we  obtain 
the Schwarzschild solution. Retaining $k$ will make $R_2^2 \not= 0$
and  will 
give rise to stresses, \\

\be
T^0_0 = T^1_1 = - \frac{k}{4 \pi r^2}.
\ee

\noindent These are precisely the stresses at large $r$ required for a global 
monopole as we shall see in the next section. Here $k$ will be related to the global  monopole  charge 
which produces neither acceleration nor tidal acceleration for radially
moving free particles. Note that the  gravitational charge density,
$4\pi \rho_c = R_{ik} u^i u^k$, continues to remain zero indicated by
vanishing of $R_{00}$.\\                               
  
\section{The Gravitational Field of a Global Monopole}

Barriola and Vilenkin [1] have discussed the
gravitational field of a global monopole formed by spontaneous
symmetry breaking of a triplet of scalar fields with a global
symmetry group $O(3)$. The Lagrangian density of the isoscalar triplet 
$\psi^a$ with $a=1,2,3$ is \\

\be
   {\cal L} = \frac{1}{2} (\partial \psi^{a})^{2} - \frac{\lambda}{4}
   (\psi^{a} \psi^{a} - \eta^{2})^{2}.
\ee

\noindent Topologically non-trivial self-supporting solutions to this
system can be found. The ansatz describing a monopole is 

\be
\psi^{a}({\bf {\bar x}}) = \eta f(r) \frac{x^{a}}{\mid {\bf {\bar x}} \mid}
\ee
   
\noindent where 

\be
x^{a} x^{a} = r^{2}
\ee

\noindent with $x^{a}$ meaning the corresponding cartesian component of
${\bf {\bar x}}$.   Outside the monopole core we must have

\be
f(r) \rightarrow 1 \ {\rm as} \ \mid {\bf {\bar x}} \mid \rightarrow
\infty 
\ee

\noindent for $V(\psi)$ to vanish asymptotically. The vanishing of
energy density is however not fast enough because 

\be
\mid \nabla \psi \mid^{2} \rightarrow \frac{\eta^{2}}{r^{2}}.
\ee   

\noindent The stress tensor of the system outside the core can be 
approximated as

\be
T^{0}_{0} = T^{1}_{1} = \frac{\eta^{2}}{r^{2}}	
\ee

\noindent and the other components vanish. These are precisely the stresses
generated (eqn.(2.5)) by keeping $k$ in (2.4).\\

Thus the most general metric for $T^0_0 = T^1_1 \neq 0$ and the remaining
stress components being zero is

\begin{equation}
ds^{2} = {(1 + 2k - \frac{2M}{r})}dt^{2} - (1 + 2k - \frac{2M}{r})^{-1} dr^2
- r^{2}d {\theta}^{2} - r^{2}sin^{2} \theta {d\phi}^{2}
\end{equation}

\noindent with $k = -4 \pi \eta^2$ and $M$ having the usual meaning of
mass of the central gravitating object [1].  
Here then central mass is that of the monopole which is
usually negligible.
However if we consider the spherically symmetric gravitational collapse of
the matter around such a monopole, we find that a black hole is formed.
The parameter $M$ then corresponds to the mass of the black hole.\\

Even when $M = 0, $ this spacetime has non-zero curvature :

\be
R_0^0 = R^1_1 = R_{01} = 0
\ee

\noindent but

\be
R^2_2 = \frac{8 \pi \eta^2}{r^2}.
\ee       

That the total energy of such solutions is divergent makes them
unrealistic except perhaps as fleeting entities in the course of
a cosmic phase transition [1]. The core mass $M$  is nevertheless 
finite. It is interesting to note
that retension of $k$ in (2.4) amounts to giving up asymptotic
flatness and that exactly generates the stresses of a global
monopole.\\

A remarkable conclusion that follows from this is that a black
hole can possess ``hair'' in the form of topological charge. 
How will this charge be detected ?
Topologically nontrivial solutions are known to have the
peculiar property that their symmetry generators are linear
combinations of spacetime symmetries and internal symmetries. In
the present case, it would mean that a motion on a sphere of
constant radius $r$ has to be accompanied by a suitable internal
symmetry transformation. An asymptotic observer looking at
particle like excitations will find that if he carries the
latter to new angular coordinate values, he has to reassign the
internal charge carried by them. For the simplest spherically
symmetric case considered, a closed trajectory in the group
space corresponding to a $2\pi$  $O(3)$ rotation will be
traversed upon orbiting the black hole. Similar conclusions
should follow for topological sectors of higher winding number,
except that the spacetime metric will not be in the simple form
corresponding to spherical symmetry.\\ 

Another interesting manifestation of the topological charge can be 
obstruction to global definition of the wavefunction of certain particle
species. For example, if the symmetry group is really the spin group 
$SU(2)$ instead of $O(3)$ (more generally, $Spin(N)$ instead of $O(N)$) 
then an isospinor species $\chi$ may exist. Now a $2\pi$ space rotation
will induce a $2\pi$ $O(3)$ rotation which will change the sign of the 
wavefunction which therefore can not be globally defined.\\

Note that the fate of the
global charge $S_{z}$ of the $O(3)$ group will be the same as
that of any other global charge such as the baryon number.

\section{Particle orbits}

In this Section we wish to investigate the effect of
global charge on the particle orbits; existence,
boundedness and stability of circular orbits, light
deflection and perihelion-shift. \\

\subsection{Circular orbits :} For the metric (3.7), we
have

\be
\bigtriangleup \dot t = E, ~ r^2 \dot \varphi = l 
\ee

\noindent where $\bigtriangleup = 1 + 2k - 2M/r, \dot t = dt/ds;~ E $ and $l$ are
specific energy and angular momentum of a test particle. Substituting (4.1)
in the metric (we have set $\theta = \pi/2 $ as usual),
\be
\dot r^2 = E^2 - \bigtriangleup \bigg({l^2 \over r^2} + 1 \bigg) = E^2 - V^2.
\ee

\noindent For existence of circular orbit, we should have both $\dot r = \ddot r = 0$.
This will give the existence threshold
\be
 r_e \geq {3M \over 1 + 2k} 
\ee

\noindent where equality refers to the photon orbit. Of course $l \geq 2 \sqrt 3
M/(1+2k)$. \\

\noindent For boundedness, we should have $E^2 \geq (1+2k)$, which leads to

\be
 r_b \geq {4M \over 1+2k} 
\ee

\noindent while stability of circular orbits will further 
require $V^{\prime \prime} \geq 0$, giving \\

\be
 r_s \geq {6M \over 1+2k} .
\ee

\noindent Thus these thresholds simply indicate scaling of 
$M \longrightarrow M/(1+2k) = M/(1 - 8 \pi \eta^2)$. \\

\subsection{Light deflection :} Following the standard calculation
[5], we write

\be
 \varphi(r) - \varphi(\infty) = \int^{\infty}_{r} 
\bigtriangleup^{-1/2} (r) \bigg[({r \over r_0})^2 {\bigtriangleup (r_0)
\over \bigtriangleup (r)} - 1 \bigg]^{-1/2} {dr \over r} 
\ee

\noindent where $r_0 $ is the radius vector of the closest approach to the 
gravitating body. \\

\noindent The deflection angle will be given by

\be
 \delta \varphi = 2|\varphi (r_0) - \varphi(\infty)| - {\pi \over
\sqrt{1+2k}}. 
\ee

\noindent Note that when $M = 0$, $\varphi$ has the period $2 \pi/\sqrt{1+2k}$
and not $2 \pi$, there is a deficit angle similar to the case of cosmic
string. From (4.6) and (4.7) we get

\be
\delta \varphi = {4M \over r_0 (1+2k)^{3/2}} \approx {4M \over r_0} 
(1 + 12 \pi \eta^2).
\ee

\subsection{Perihelion-shift :} Here the corresponding relation [5]
is 

\begin{eqnarray}
\varphi(r) - \varphi(r_{-}) &=& \int^r_{r_{-}} \bigg[{r^2_- (
\bigtriangleup^{-1}(r) - \bigtriangleup^{-1} (r_-)) - r^2_+ (\bigtriangleup^{-1}(r)
- \bigtriangleup^{-1} (r_+)) \over r^2_+ r^2_- (\bigtriangleup^{-1} (r_+) -
\bigtriangleup^{-1} (r_-))} \nonumber \\
&  &- {1 \over r^2} \bigg]^{-1/2} {\bigtriangleup^{-1/2} \over r^2} dr  
\end{eqnarray}

\noindent where $r_{\pm}$ refer to aphelion and perihelion points. The perihelion
shift is given by

\be
 \delta \varphi = 2 |\varphi(r_+) - \varphi (r_-)| - {2 \pi \over \sqrt{1+2k}} \ee

\noindent and we thus obtain

\be \delta \varphi = {6 \pi M \over L(1+2k)^3/2} \approx {6 \pi M \over L}
(1 + 12 \pi \eta^2) 
\ee

\noindent where

$$L = {1 \over 2} \bigg({1 \over r_+} + {1 \over r_-}\bigg). $$

Thus effect of global charge is to scale the Schwarzschild values by
$(1 - 8 \pi \eta^2)^{-1} $ for existence, boundedness and stability thresholds of
circular orbits and by $(1 - 8 \pi \eta^2)^{-3/2}$ for light deflection and
perihelion-shift. It means a slight enhancement in
the Schwarzschild values.\\
              
\section{Semiclassical Effects}       

\par
Next we investigate the question of Hawking radiation in this
background metric. Since the thermal nature of the radiation is
tied to the existence of the horizon, we do not expect this to
change. However the temperature parameter may carry a signature
of the global monopole.

Consider the equation satisfied by a massless scalar quantum
field $\psi$

\be
\Box \psi = 0 .
\ee

\noindent The advanced and retarded coordinates for this spacetime are 

\begin{equation}
v = t + r^{*}
\end{equation}
\begin{equation}
u = t -r^{*}
\end{equation}

\noindent where $r^{*}$ is the tortoise coordinate defined in this case as

\begin{equation}
dr^{*} = \frac{r}{1 + 2k - \frac{2M}{r}} .
\end{equation}

\noindent The wave equation can be separated in ($t,r,\theta,\varphi$)
coordinates. The mode functions relevant to this spacetime are of the form

\begin{equation}
\frac{R_{\omega l}(r)}{r} Y_{lm}(\theta ,\varphi) e^{-i\omega t}
\end{equation}
$Y_{lm}$ is a spherical harmonic and $R_{\omega l}$ satisfies the 
equation

\begin{equation}
\frac{d^{2}R_{\omega l}}{dr^{*2}} + (\omega^{2} - [l(l+1)r^{-2}
+ 2Mr^{-3}][1 - 2Mr^{-1}])R_{\omega l} = 0.
\end{equation}
Because of the `potential' term in square brackets, the standard incoming
waves will partially scatter back off the gravitational field to become a
superposition of incoming and outgoing waves. However, the effective 
potential vanishes as $r^{*} \rightarrow \pm \infty$ (i.e.for $r
\rightarrow 2M,\infty$). Thus, in those regions the mode functions will be of 
the form 

\begin{equation}
(\frac{1}{r} e^{-i \omega u} Y_{lm})   \  {\rm{and}}  \  
(\frac{1}{r} e^{-i \omega v} Y_{lm}).
\end{equation}
The surface gravity of the black hole formed as the result of
gravitational collapse is
       
\be
\kappa = \frac{(1+2k)^{2}}{4M} = \frac{(1 - 8 \pi \eta^{2})^2}{4M}.
\ee
\noindent
This redefinition of the surface gravity reduces the calculation
to that of the Schwarzschild case. With this redefinition, the
Penrose diagram of the present system is identical to that of
the Schwarzschild case.

\noindent As in the Schwarzschild case, assuming there are no bound states
at late times, the ingoing and outgoing modes form a complete
basis in terms of which the field at late times can be expanded.
Then an outgoing mode of the form $r^{-1} e^{-i \omega u}
Y_{lm}$ at late times can be traced back to early times in the
same fashion as in the Schwarzschild case. At early times, the
ray, moving along constant phase $v$ lines will have the form
$e^{-i\omega u(v)}$. The affine parameter on the past horizon can
be chosen as 

\be
\lambda = - c e^{-\kappa u}.
\ee
\noindent Then the function

\be
u(v) = - \frac{1}{\kappa} \ln[\frac{v_{0} - v}{{\rm const}}].
\ee
\noindent
This can be Bogolubov transformed in the same manner as the
Schwarzschild case to give the Planckian spectrum 

\begin{eqnarray}
N_{\beta} &=& \frac{1}{e^{2\pi \omega/\kappa} - 1}\\
  &=& \frac{1}{e^{8\pi M \omega/(1+2k)^{2}} - 1}.
\end{eqnarray}

\noindent
Thus the thermal spectrum is recovered but the temperature is
now $(1 - 8 \pi \eta^2)^2/8 \pi M$. 

Consider next the quanta of the
$\psi$ field responsible for the monopole.
The equations satisfied by 

\be
\tilde{\psi^{a}} = \psi^{a} - \psi^{a}_{bg} 
\ee
are, after linearisation,

\be
\Box \tilde{\psi^{a}} + 2 \lambda \eta^{2} \psi^{a}_{bg} (\sum_{b} \psi^{b}_{bg} \tilde{\psi^{b}}) = 0.
\ee
Asymptotically the equation becomes 

\be
\Box \tilde{\psi^{a}} + 2 \lambda \eta^{2} 
\frac{x^{a}}{r^{2}} (\sum_{b} x^{b} \tilde{\psi^{b}}) = 0.
\ee
Consider the $a=3$ mode and the asymptotic $z$ direction, 
the other two ($x$ and $y$) remaining finite. Then the equations become

\be
\Box \tilde{\psi^{3}} + 2 \lambda \eta^{2} \tilde{\psi^{3}} = 0
\ee
\be
\Box \tilde{\psi^{1}} = 0
\ee
\be
\Box \tilde{\psi^{2}} = 0 .
\ee
This corresponds to the radial mode being massive but the two transverse 
modes being massless. Thus the preceding analysis will apply to
the two transverse modes as well. We see that an imprint of the
topological charge of the black hole is left on the Hawking
radiation as well, namely under angular displacement, different
components of the $\psi$ triplet will be detected.\\

Thus the fate of our black hole is similar to that of the magnetically
charged black holes [6] . While the question
of its ultimate fate cannot be settled within the semiclassical
approximation, the answer will be the same as for a
Schwarzschild black hole. If it can in fact evaporate
completely, and does so, then it leaves behind a monopole identical to the
original one before collapse.\\

In concluding this section, we note that the persistence of the
thermal radiation is not surprising. It is known from work in
axiomatic field theory [7] that when horizons exist, a
unique nontrivial automorphism exists for the ground state. The
automorphism works out to be periodicity in time, leading to the
Kubo-Martin-Schwinger boundary condition on the Green function.
\\
\section{Discussion}

The main purpose of this investigation was to study the effect of
global monopole charge on particle orbits and the Hawking radiation.
It turns out that existence, boundedness and stability threshold
for circular orbits scale up by $(1 - 8 \pi \eta^2)^{-1}$, perihelion
shift and light bending by $(1 - 8 \pi \eta^2)^{-3/2}$, while the
Hawking temperature scales down by $(1 - 8 \pi \eta^2)^2$ the
Schwarzschild values. This is how the global monopole charge $\eta$
affects the particle orbits and the Hawking radiation thermal spectrum.
It may be noted that the event horizon is given by $r = 2 M(1 - 8 \pi
\eta^2)^{-1}$, while the red-shifted proper acceleration remains 
unchanged as $M/r^2$. It is the scaling up of the horizon that leads
to scaling down of the surface gravity (temperature) of the hole.\\

The spacetime (3.7) with $M=0$ can be looked upon as ``minimally''
curved for it has zero gravitational charge $(R_{ik} u^i u^k = 0)$
which leads to vanishing of acceleration as well as tidal acceleration
for radial motion. Its curvature only manifests in producing tidal
acceleration for transverse motion. \\

\noindent{\bf Acknowledgement :}  KN thanks Jawaharlal Nehru Centre
for Advanced Research, Bangalore for award of a summer fellowship and
IUCAA for providing the facilities.

\end{document}